# Progressive Compression of 3D Objects with an Adaptive Quantization

Zeineb ABDERRAHIM[1], Elhem TECHINI[1], Mohamed Salim BOUHLEL[1]

[1] University of Sfax, High School Engineering of Sfax (ENIS),
Research Unit Sciences and Technologies of image and Telecommunications (SETIT)

**Abstract**

This paper presents a new progressive compression method for triangular meshes. This method, in fact, is based on a schema of irregular multi-resolution analysis and is centered on the optimization of the rate-distortion trade-off. The quantization precision is adapted to each vertex during the encoding / decoding process to optimize the rate-distortion compromise. The Optimization of the treated mesh geometry improves the approximation quality and the compression ratio at each level of resolution. The experimental results show that the proposed algorithm gives competitive results compared to the previous works dealing with the rate-distortion compromise.

## 1. Introduction

Thanks to the increase of computer performance (computing capacity, memory and graphics) and advanced Internet networks, the three-dimensional objects are increasingly becoming used in our daily lives and in many fields including computer-aided design (CAD), scientific simulation, augmented reality, virtual visits, medicine, video games or movies.

These 3D objects are usually presented in the form of a triangular surface mesh due to their efficiency and simplicity: the geometric structure is mainly composed of geometry and connectivity. In other words, geometry presents the vertex positions, but connectivity presents the way how the vertices are connected together to form triangles.

Although all this information is a powerful tool for representing and modeling 3D objects, even very complex ones, it also represents a significant amount of data to be kept for fine and detailed modeling. Indeed, these 3D objects increase in complexity and detail what requires, despite the expansion of networks, effective compression techniques to allow not only a compact representation of the object, but also a reduction of the storage size and transmission time of the 3D object on the network.

In this context, the progressive compression methods are very effective because they offer the ability to visualize, compress and progressively transmit the object at different levels of detail.

Therefore, these methods allow the adaptation transfer of these objects to different client resources (rate of networks, visualization capacity of the terminal).

The main challenge of these approaches, however, is to optimize the rate-distortion compromise in order to obtain an approximation of the final mesh that can be as faithful as possible to the original one throughout the refinement process. As a matter of fact, quantization is an important factor in optimizing geometry. Generally, there are techniques that apply an initial quantization and a fixed precision to vertices of mesh. Other techniques use an adaptation of quantization on each resolution level. We observe, also, the strong relationship between the quantization precision and the complexity of the intermediate meshes. Indeed, our work is aimed to improve a progressive compression method based on multi-resolution analysis which focuses on irregular meshes using an adaptation of quantization precision for each mesh vertex to optimize rate-distortion compromise.

This article consists of three parts. In the first part, we present the principal compression of the 3D mesh and the existing rate-distortion optimization. In the second part, we present the basic algorithm used in our approach then we describe our method of improvement of the rate - distortion. The last part is devoted to show the results obtained by our proposal and to compare them with some methods presented in the previous section. Finally, this article is ended up by a conclusion.

## 2. State of the Art

We can classify the compression algorithms of 3D meshes in two main categories. The first category includes mono-resolution compression techniques that allow lossless compression and require full decoding before visualizing the object. The second entails the progressive compression techniques that allow compression and transmission of various levels of resolution. For a thorough study the reader may refer to [1] [2].

The mono-resolution compression techniques are the first techniques that were proposed, they are meant to encode the network with a minimum number of bits by eliminating the existing redundancy in the original representation of connectivity. And the global form of the original model is accessible at the end of the transmission.
The first work is introduced by Deering [3] who suggested a coding by band of generalized triangle. This technique requires an appropriate division of the mesh in order to achieve an efficient compression. Many works result from this approach [4]. Rossignac proposed an algorithm called "edgebreaker" guaranteeing maximum coding cost of 2 bits per face and 3.67 bits per vertex but this time using an alphabet with 5 codes instead of a binary stream to describe the bands of the triangle. This method is interesting only for too irregular meshes.

A further work based on coding by valence is then introduced by [5] who proposed an algorithm defined for the oriented mesh manifold. The valence of the vertices can be encoded efficiently by entropic coding. Thus, it is mainly effective for regular meshes whose topological cost tends towards zero (0.2 bpv). The geometry is encoded by prediction using the parallelogram rule.
This approach was improved by Alliez and Derbrun in 2001 [6]. They proved that the upper bound of their encoder is 3.24 bits per vertex for large irregular meshes. More recently, Isenburg also proposed an alternative [7], the polygonal surface meshes based on ASCII encoding.

In contrast, the progressive compression methods rested on a coarse mesh and refinement information gradually transmitted to obtain different levels of mono-resolution. Unlike the mono-resolution methods, they offer the possibility of accessing the intermediate reconstructions of a 3D object during its transmission.
The first progressive compression method was introduced by Hoppe [8] which consists in iteratively applying an operation edge contraction "edge collapse" during the encoding to generate a coarse mesh. The reconstruction is done by a reverse operation "vertex split" while decoding. The total cost is non-linear (n.log (n)); it is reserved for meshes of low complexity.

Cohen-Or et al [9] propose a simplification algorithm based on the technique of color patches and the successive withdrawals of vertices followed by deterministic re-triangulation. These patches are colored by using 2 or 4 colors to allow the decoder to properly insert vertices in the reconstruction based on the color patches. This algorithm compresses connectivity with a cost of 6 bits / vertex.
The previous algorithm was recently improved by Alliez and Desbrun [10] who introduced an algorithm based on the progressive deletion of independent vertices with a re-triangulation step under the constraint of maintaining the vertex degrees around 6. It has an average coding cost of 1.85 bits / face. The conservation of the regularity ensures an efficient connectivity coding and the connectivity coding cost is around 3.7 bits / vertex. All these approaches are directed by the connectivity as they give priority to connectivity coding. For this, the most recent progressive encoders are not only motivated by the refinement of connectivity, but also by the geometrical measurements of the rebuilt mesh.

Gandoin and Devillers [11] [12] propose a compression algorithm which is no longer driven by connectivity but by the geometry based on the kd-tree subdivision. The main objective in their approach is to achieve continuity between the coarse approximations and the finest ones of the object. The connectivity is encoded by encoding all the change in the cell subdivision. In terms of compression rate, this algorithm has better results than the algorithms guided by connectivity.
This method is improved by Peng and Kuo [13] by replacing kd-tree with Oc-tree data structure. The Oc-tree cells are refined in order of priority, where subdivisions of cells with the greatest improvement of distortion are executed the first. By optimizing the coding of connectivity and geometry through effective predictions, this approach gives good results for lossless compression and improved from 10 to 20% compared to [11].

Unlike the previous approaches, the spectral approach introduced by Karni and Gotsman [14] proposed a spectral decomposition by projecting the mesh geometry on a set of eigenvector from the diagonalization of the discrete Laplacian operator. This approach is generally effective for low compression rate of smooth geometry models. This results in average costs of coding around 4.5 bits / vertex. For this, these methods do not offer a progressive geometry; connectivity remains unchanged during the transmission. Indeed, most works are directed towards

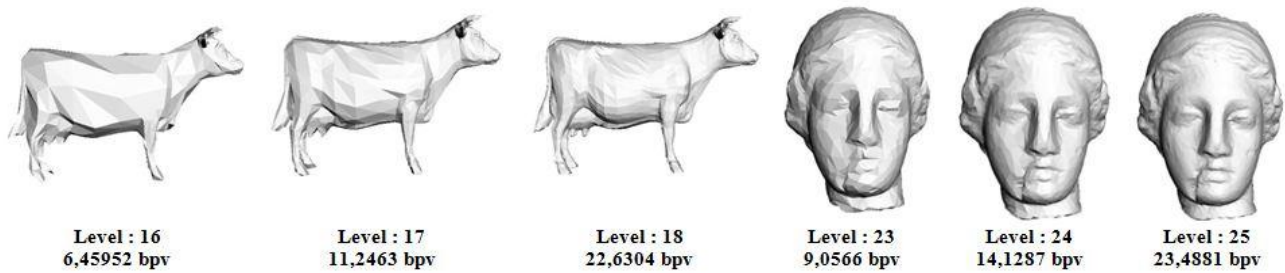

Fig 1. Compression results of our approach for some meshes at different rates.

multi-resolution analysis which allows having both geometric and topological scalability.

At last, we can distinguish those which were focused on semi-regular meshes based on remeshing and those which were adapted to the topological structure of the irregular mesh.

The pioneer progressive compression method of the semi-regular mesh was proposed in 2000 by Khodakovsky et al [15]. It consists in remeshing an arbitrary mesh M into a semi-regular mesh by using the MAPS algorithm (Multi-resolution Adaptive Parameterization of Surfaces) [16]. After this remeshing step, Khodakovsky et al choose to apply a transformed semi-regular wavelet based on the loop interpolating filter.

A recent method has been proposed by Payan et al [17] which is the compression algorithm with loss for dense triangular meshes incorporating a binary-allocation. This method is based on a stage of remeshing using a normal remesher and one transformed into a non lifted Butterfly wavelet.

For this, other existing techniques directly consider the irregular structure of the mesh such as the Wavemesh [18] which proposed a new approach guided by connectivity and extended the wavelet decomposition scheme on an irregular mesh so as to generate approximations of better quality than that based on geometry. The connectivity coding cost is 2 bits / vertex on average.

In 2009, Valette et al [19] proposed a new lossless progressive compression algorithm of mesh based on a metric "Incremental Parametric Refinement" named "IPR", where connectivity is not controlled in a first step, giving visually pleasant meshes to each level of resolution while saving connectivity information compared to the previous approaches. This approach shows its effectiveness in terms of rate-distortion, especially at low rate.

To optimize the rate-distortion compromise for meshes during transmission, optimization techniques are also used. The algorithm [17] optimizes the accuracy of quantization of the wavelet coefficients of each sub-frequency band so as to minimize the geometric error for a given rate.

Thus, the binary allocation used to control the visual quality of the reconstructed mesh while encoding the geometry to optimize the rate-distortion compromises [20] regardless of the desired compression rate.

In this context, [19] presents an approach that adapts the quantization precision of each vertex based on the local geometric configuration. This method gives good results especially at low rate.

More recently, [21] [22] presents a new method for optimizing the rate-distortion based on the adaptation of the quantization precision of geometry and color for each intermediate mesh. The used adaptation, which is performed at each iteration and is chosen from the decimation operation and the operation of decreasing the quantization precision, highly improves the rate-distortion trade-off.

## 3. Contribution

### 3.1 Overview

Generally, the algorithms run by connectivity produce intermediate meshes of higher quality than the algorithms guided by geometry. Thus, these algorithms are more efficient than those based on connectivity for the rate-distortion compromise.

Hence, our method is founded on the Wavemesh algorithm [18], which is an algorithm guided by connectivity. This approach, indeed, is able to directly adapt to the topological structure of irregular meshes offering new irregular subdivision schemes ("Fig. 2"). The goal is to overcome the difficulty of connectivity subdivision imposed by the method of Lounsbery and maintain the geometry and connectivity of the approximations as close as possible to the original mesh. Additionally, this approach avoids the remeshing process that overcomes the difficulties of its implementation for some mesh from medical imaging and that presents an additional gain. The results of the Wavemesh indicate its superiority and its performance compared to other approaches in terms of lossless progressive compression.

The quantization precision is an important factor used to optimize the rate distortion. Optimization is often neglected in most researches on progressive compression. Thus, most of the existing methods use a 12 bit-scalar-uniform quantization and some approach gets around an adaptation of quantization on each level of detail.

In addition, we notice that the quantization precision is larger than necessary for intermediate meshes at low resolution for algorithms guided by connectivity that retain high precision of the initial quantization throughout the simplification [19].

Indeed, to optimize the rate-distortion compromise, our progressive compression method is based on the adaptive quantization algorithm IPR [19] that uses a technique of adaptive quantization on each vertex.

In the next section, we describe our proposed encoding and adaptive quantization of geometry.

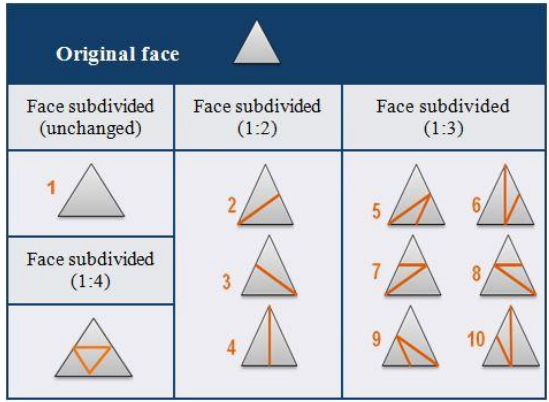

Fig 2. Possible cases of subdivision of the Wavemesh method [18].

### 3.2 Wavelet transform

The algorithm starts with the simplification of the original mesh by inverting the irregular subdivision schemes which is described in detail in [23].

A mesh $M_j$ is simplified by an inverse irregular subdivision scheme where each face can be divided into four, three or two faces, or remain unchanged. Simplification is repeated until the original mesh $M_j$ cannot be simplified.

Moreover, this simplification is implemented in two ways: The first way allows to simplify the mesh regardless of the mesh geometry so that the connectivity is the best simplified while the second way uses a Wavelet Geometric Criterion (WGC) which tends to improve the rate-distortion efficiency of the compression algorithm.

When the simplification is over, we can reconstruct a hierarchical relationship between the original mesh $M_j$ and the simplified one $M_{j-1}$.

Thus, the geometry of $M_{j-1}$ can be calculated by the approximation $M_j$ by applying the wavelet decomposition with the two analysis filters $A^j$ and $B^j$. Indeed, they are represented in the following matrix form:

$$C^{j-1} = A^j . C^j \quad (1)$$
$$D^{j-1} = B^j . C^j \quad (2)$$

With $C^j$ is the matrix $N(j)*3$ giving the coordinates of each vertex of the mesh at the resolution j.
$C^{j-1}$ is the geometrical approximation of the mesh and $D^{j-1}$ the detail coefficients or wavelet coefficients which is required to reconstruct $C^j$ starting from $C^{j-1}$
The reconstruction of the high resolution mesh geometry $C^j$ is provided by both synthesis filters $P^j$ and $Q^j$ taking into account the geometry of coarse mesh (low resolution) $C^{j-1}$ and the wavelet coefficients $D^{j-1}$

Thus, in order to obtain the best approximation, the lifting scheme is used. The wavelet decomposition, which is used to make the approximations, transforms the vertex coordinates into a wavelet coefficient which will undergo a quantization as explained below.

### 3.3 Adaptive quantization and coding

The quantization of the vertex coordinates plays an important role in the efficient compression of meshes. As such, in the compression approaches based on connectivity, the Wavemesh usually encodes and transmits the vertex coordinates with a constant quantization, while the compression schemes Octree and Kd-tree do not provide a sound balance between the number of mesh vertices (the complexity of the mesh) and the quantization precision of the vertex coordinates.

The quantization algorithm applied at this stage consists in quantifying each vertex with a specific quantization precision. Indeed, the quantization precision is the number of bits used for quantization and it is determined according to the following procedure:

The first step is to calculate for each vertex $S_i$ its closest neighbor, let us say the vertex $S_j$. Then, the quantization of coordinates of these vertices $S_i$ and $S_j$ using the following equation:

$$\tilde{c}_i = \left\lfloor 2^{Q_i - Q_m} c_i \right\rfloor \quad (4)$$

Table 1.1 Comparison of the total rate of compression by bit per vertex.

| Model | V | Q | AD01[15] | VP04[16] | PK05[19] | IPR09[32] | LE11[33] | Our approach |
|---|---|---|---|---|---|---|---|---|
| *Fandisk* | 6475 | 10 | 17.4 | 13.5 | 13.3 | 12.4 | 16.1 | **12.11** |
| *Bones* | 2154 | 12 | - | 49.6 | - | - | - | **42.6** |
| *Venushead* | 8268 | 12 | - | 26.7 | - | - | - | **23.5** |
| *Cow* | 2903 | 12 | - | 25.7 | - | - | - | **22.6** |
| *Venusbody* | 711 | 12 | - | 31.6 | - | - | - | **23.3** |

$Q_i$ and $Q_m$ respectively represent the initial quantization precision of $S_i$ and the value of the maximum precision such as $3 < Q_i < Q_m$ with $Q_m = 12$ while $C_i$ represents the coordinate of a given vertex $S_i$.

To determine whether the quantization precision associated with the vertex $S_i$ is sufficient or not, the quantization algorithms have to compute the squared distance between the two vertices $S_i$ and $S_j$ in its quantified version. This is done by the following formula:

$$D_i = \left\| \tilde{c}_i - \tilde{c}_j \right\|^2 \quad (5)$$

The sufficient precision is then achieved only if the value of the distance $D_i$ exceeds a threshold $Q_t$. According to the experiment that we carried out, we choose not to keep the threshold value proposed in the basic algorithm which is equal to 600 but to use the value which seems to be the most appropriate and equal to 200. In the case that the quantization precision is not sufficient, we increment it with one bit and we keep iterating until the result of quantization precision is sufficient.

This technique of adaptive quantization can be summarized by the following algorithm:

```
Algorithm : Quantization algorithm
Begin
  For ( i = 0  to Number of vertices increment 1 ) do
    Calculate, Sj the nearest neighbor of Si
    Initialization of Qi
    Repeat
      Quantify the coordinates of Si and Sj
      Calculate distance between vertices Si  and Sj (quantized version)
      if ( distance < 200) then
        Recover coordinates before quantization of Si and Sj
        Increase by a bit the value of quantization Qi
      end
    Until ( distance <=200)
    Recover coordinates of vertex Sj before quantization
  End for
End
```

Fig 3. Quantization algorithm of our approach

The quantified wavelet coefficients are compressed by means of adaptive arithmetic encoder.

## 4. Experimental results

### 4.1 Compression ratio

Table 1.1 shows the total compression rates of our approach applied to various models and compared with some results from [18], [19], [10] and [22]. The index values indicated with "-" are not tested.

All models were quantified in 12-bit by coordinates, with the exception of the Fandisk model (quantified in 10 bits). The observation results provided in Table 1.1 show that our method can significantly reduce the total rate of compression. The gain shown is from 26% to 33%. We also notice that for the Fandisk model our algorithm has a gain similar to the IPR method, which is better than the AD method and LE. Our approach improves the compression rate compared to the Wavemesh method ranging from 1.39 to 8.3 bits / vertex.

Table 1.2 lists the second compression rate of geometry of our proposal compared with that of Valette and Prost [18] and AD [10].

On both models Fandisk and Nefertiti, we presented two results. Indeed, the first result indicates the implementation by considering the condition WGC and the second result is without WGC.

When comparing the results obtained with and without WGC, we notice that the WGC improves the compression rate of geometry for the Fandisk model about 0.7 bit / vertex. On the other hand, for Nefertiti model, the result is less effective because the WGC does not work properly on smooth meshes.

Tableau 1.2 Comparison of the compression-rates of geometries

| Model | Vertices | Wavemesh (12 bits) | Our approach | AD (10 bits) |
|---|---|---|---|---|
| *Fandisk+WGC* | 6475 | 14.70 | 8.84 | 12.34 |
| *Fandisk* | 6475 | 15.71 | 9.55 | - |
| *Nefertiti+WGC* | 299 | 30.65 | 22.04 | 11.88 |
| *Nefertiti* | 299 | 30.52 | 21.69 | - |
| *Venusbody* | 711 | 26.87 | 18.66 | - |
| *bones* | 2154 | 29.19 | 25.50 | - |

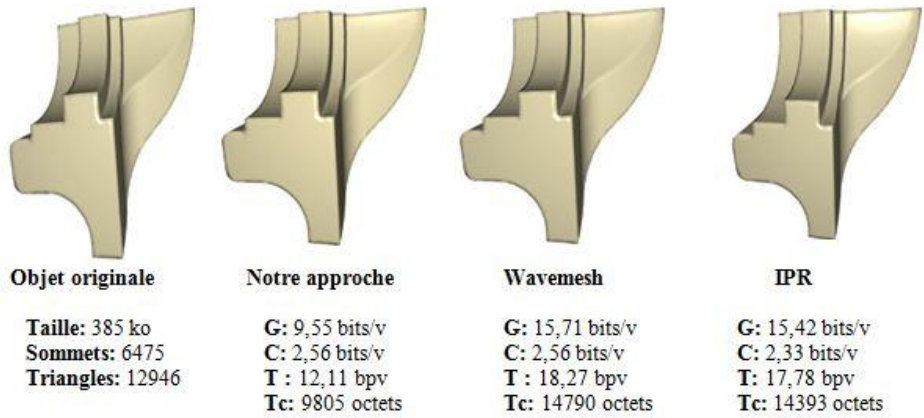

Fig 3. Results of model Fandisk obtained by different approaches: our algorithm Wavemesh [18] and IPR [19].With (G: Geometry C: Connectivity; T: total Data (C+G) ; Tc: File size).

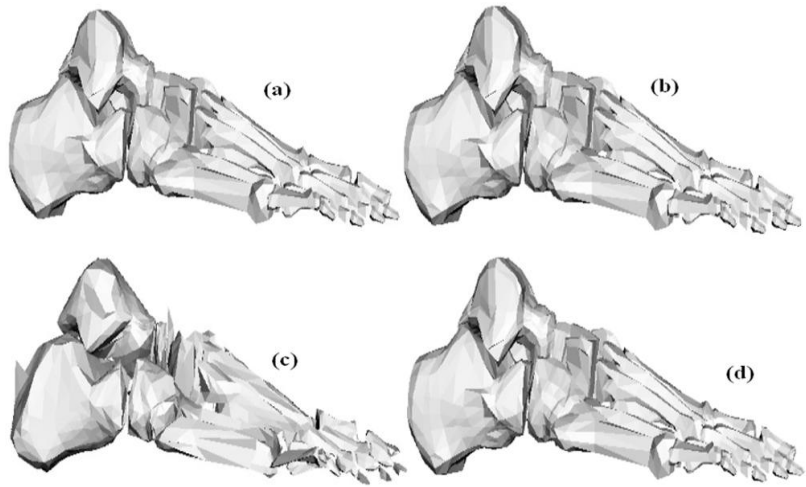

Fig 4. Compression results for the object Bones (a) of : the basic algorithm Wavemesh[18] (b), algorithme described in [24] (c) and our approach (d).

proposed compression scheme leads to better geometric compression results than the one presented by the basic algorithm. It provides a compression gain ranging from 12% to 70%. We note that at low rate the performance of our method outperforms the Wavemesh with and without the condition WGC.

4.2 Comparison of visual performance

Figure 1 shows the compression results of our approach and the wavemesh approach for some meshes at different rates. We note that our approach retains a good visual quality all along the mesh refinement process by minimizing the size of the file.

Figure 3 presents the visual extracts of the highest level of Fandisk mesh resolution respectively provided by our algorithm, the wavemesh and IPR. Compared to these different methods, our approach gives a very good visual result since the size of the compressed file is minimized.

In Figure 4 we compare the visual appearance to be provided by the bones basic method, wavemesh, and the precision progressive method proposed by [24]. The comparison of these images shows a visual similarity between the results provided by our algorithm and the basic method and the superior visual quality of our result with respect to the method of [24].

## 4.3 Rate-distorsion curves

Figure 6 and Figure 7 respectively show the rate-distortion curve of Nefertiti and Mannequin models. Both models are quantified to 12 bits for the Wavemesh algorithm and to a quantization precision between 4 and 12 bits for our proposal.

In these figures, the vertical axe indicates the amount of distortion which is the maximum RMS (root mean square) distance normalized by the diagonal of the bounding box. The RMS value is measured by the Metro tool [25] while the horizontal axis shows the rate in bit per vertex.

These curves show that our algorithm optimizes the rate-distortion especially at low rate.
However, for rate-rates below 10 bps our approach leads to better results. On the other hand, for rates higher than 10 bps, the visual quality of the generated meshes becomes almost constant and the encoders of the Wavemesh and IPR provide slightly higher rate-distortion reports. Indeed, this approach proves its high efficiency at low rate rates.

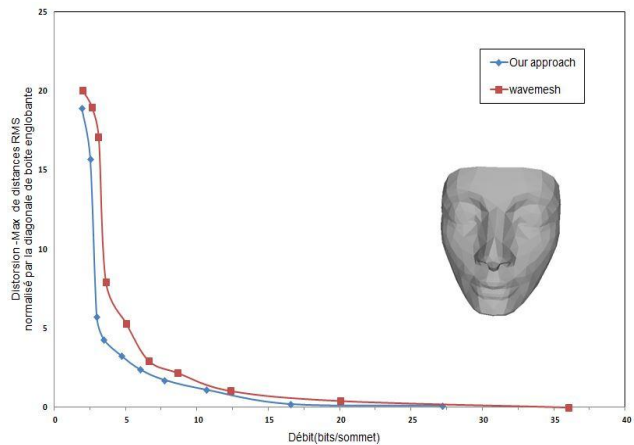

Fig 6. Rate-distortion curve of Nefertiti.

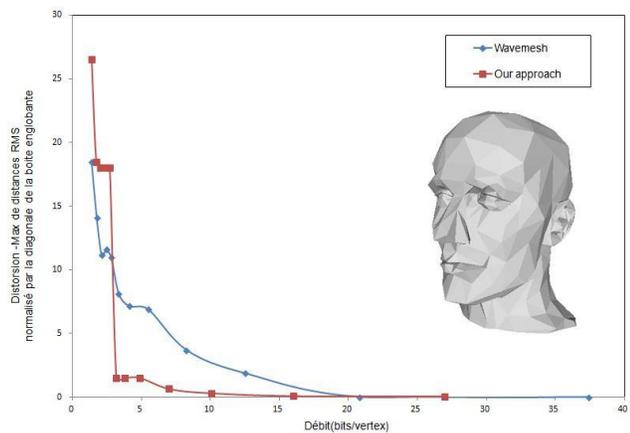

Fig 7. Rate-distortion curve of Mannequin.

## 4. Conclusion

In this paper, we proposed a new method for progressive mesh compression based on the adaptation of the quantization precision of geometry which allows to improve the rate-distortion, especially at low rate.
The results show that this method of adapting quantization yields better results than the classic algorithms in terms of rate-distortion and very satisfactory visual results.

This work can be improved through the integration of different attributes such as color and normal. Thus, our perspective is to find a theoretical value for the measured threshold for determining the quantization precision so as to ensure a balance with the density of vertices.

### Acknowledgments

We thank the anonymous reviewers for their valuable comments which helped improving the quality of the paper. We also thank Sébastien Valette for fruitful discussions on compression method "wavemesh". Our implementation is based on the Visualization ToolKit (www.vtk.org).

### References

[1] PENGJ., KIM C.-S., KUOC.-C. J.: Technologies for 3D mesh compression: A survey.Journal of Visual Communication and Image Representation 16, 6 (2005), 688–733.

[2] ALLIEZP., GOTSMANC.: Recent advances in compression of 3d meshes. InProc. of the Symp. on Multiresolution in Geometric Modeling(Sep 2003).

[3] Michael DEERING. Geometry compression. In SIGGRAPH, pages 1320, 1995.

[4] ROSSIGNACJ.: Edgebreaker: Connectivity compression for triangle meshes. IEEE Trans. Visualization and Comp.Graphics 5, 1 (1999), 47–61.

[5] C.Touma and C.Gotsman, "Triangle mesh compression", pp. 26–34, Graphics Interface,1998.

[6] P. Alliez and M. Desbrun, "Valence-Driven Connectivity Encoding for 3D Meshes", EUROGRAPHICS 2001, Volume 20, No. 3, 2001, 10p.

[7] M. Isenburg and J. Sonoeyink, "Coding with ascii : compact, yet text-based 3D content", First International Symposium on 3D Data Processing Visualisation and Transmission, pp.609-616, 2002.

[8] Hugues HOPPE.Progressive Meshes. In SIGGRAPH, pages 99108, 1996.

[9] D. Cohen-Or, D. Levin, and O. Remez, "Progressive compression of arbitrary triangular meshes," in IEEE Visualization 99, 1999, pp. 67–72.

[10] Pierre ALLIEZ et Mathieu DESBRUN. Progressive compression for lossless transmission of triangle meshes. In SIGGRAPH'01, pages 195202, 2001.

[11] O.Devillers and PM.Gandoin, "Comression interactive de maillage triangulaire arbitraires".Rapport de recherche ,INRIA Sophia Antipolis,2001,28p.

[12] Olivier DEVILLERS and Pierre-Marie GANDOIN.Progressive Lossless Compression of Arbitrary Simplicial Complexes. ACM Transactions on Graphics, vol. 21, no. Siggraph'2002 Conference pro ceedings, pages 372379, 2002

[13] J.Peng, C.Kuo -C. J.: "Geometry-guided progressive lossless 3d mesh coding with octree (ot) decomposition". ACM Trans. Graph. 24, 3 (2005), 609–616. 2, 3, 4, 8, 9 .

[14] Z. Karni and C. Gotsman, "Spectral Compression of Mesh Geometry", ACM Siggraph Conference Proceedings, pp. 279–286,2000.

[15] A. Khodakovsky, P. Schröder, W. Sweldens, "Progressive Geometry compression", International Conference on Computer Graphics and Interactive Techniques, SIGGRAPH 2000, 2000, pp. 271-278.

[16] A. W. F. Lee, W. Sweldens, P. Schröder, L. Cowsar and D. Dobkin, "MAPS : Multiresolution Adaptive Parametrization on Surfaces", International Conference on Computer Graphics and Interactive Techniques, SIGGRAPH'98, Orlando, Florida, USA, July 1998.

[17] F. Payan and M. Antonini, "An efficient bit allocation for compressing normal meshes with an error-driven quantization", Computer Aided Geometric Design, 2005, vol.22,no 5, p.466–486.

[18] Sébastien VALETTE et Rémy PROST. Wavelet-Based Progressive Compression Scheme for Triangle Meshes : Wavemesh. IEEE Trans. Vis. Comput. Graph., vol. 10, no. 2, pag es 123-129, 2004

[19] Sébastien VALETTE, Raphaëlle CHAINE et Rémy PROST.Progressive Lossless Mesh Compression Vi a Incremental Pa-rametric Renement. Computer Graphics Forum (Pro ceedings of Sym-p osium on Ge ometr Pro cessing 2009), no. 5, page 13011310, Jul 2009.

[20] F. Payan, Optimisation de compromis débit- distorsion pour la compression géométrique de maillages surfaciques triangulaires ,Thèse de doctorat, STIC de Nice_Sophia Antipolis, 2004.160p.

[21] Ho. Lee, "Compression progressive et tatouage conjoint de maillages surfaciques avec attributs de couleur", thése doctorat, Université Claude Bernard ,Lyon 1 21juin 2011,160p.

[22] Ho. Lee, G. Lavoue, F. Dupont, "Rate-distortion optimization for progressive compression of 3D mesh with color attributes", The Visual Computer 28(2): 137-153 (2012).

[23] S. Valette and R. Prost, "Wavelet based multiresolution analysis of irregular surface meshes," IEEE Transactions on Visualization and Computer Graphics, to appear, available upon request, 2003.

[24] Valette, Sébastien. Prost, Rémy. Gouillard, Alexandre. "Compression of 3D triangular mesh with progressive precision" .Computers & Graphics 2004 vol 28 no 1, p.35-42.

[25] Paolo CIGNONI, Claudio ROCCH INI et Rob erto SCOPI-GNO. Metro : Measuring Error on Simplied Surfaces. Computer Graphics Forum, vol. 17, no. 2, pages 167174, 1998.

**Zeineb Abderrahim** is currently a PhD student at Sfax University, Tunisia. She was born in Gabes Tunisia) in October 1985. She received the diploma in Computer Science and multimedia in 2008 and the master's degree in computer Science in 2010 from the higher Institute of Computer Science and Multimedia of Gabes (ISIMG). She was a member in Unit Sciences and Technologies of image and Telecommunications (SETIT). Her research interests are 3D Image processing, wavelets, progressive compression and multiresolution analysis.